\documentclass[aps,prl,citesort,twocolumn]{revtex4-1}

\usepackage{graphicx}
\usepackage{amssymb,amsmath}
\usepackage{color}
\usepackage{verbatim}

\usepackage{lineno}

\begin{document}


\title{A test of electric charge conservation with Borexino}

\author{
M.~Agostini$^{s,u}$,
S.~Appel$^s$,
G.~Bellini$^k$,
J.~Benziger$^o$,
D.~Bick$^d$,
G.~Bonfini$^j$,
D.~Bravo$^q$,
B.~Caccianiga$^k$,
F.~Calaprice$^{n,u}$,
A.~Caminata$^c$,
P.~Cavalcante$^j$,
A.~Chepurnov$^t$,
D.~D'Angelo$^k$,
S.~Davini$^u$,
A.~Derbin$^m$,
L.~Di~Noto$^c$,
I.~Drachnev$^u$,
A.~Empl$^v$,
A.~Etenko$^h$,
K.~Fomenko$^{b}$,
D.~Franco$^{a}$,
F.~Gabriele$^j$,
C.~Galbiati$^{n,k}$,
C.~Ghiano$^c$,
M.~Giammarchi$^k$,
M.~Goeger-Neff$^s$,
A.~Goretti$^{j,n}$,
M.~Gromov$^{t,za}$,
C.~Hagner$^d$,
E.~Hungerford$^v$,
Aldo~Ianni$^{j,zb}$,
Andrea~Ianni$^{n,j}$,
K.~Jedrzejczak$^{g}$,
M.~Kaiser$^d$,
V.~Kobychev$^f$,
D.~Korablev$^{b}$,
G.~Korga$^j$,
D.~Kryn$^{a}$,
M.~Laubenstein$^j$,
B.~Lehnert$^w$,
E.~Litvinovich$^{h,i}$,
F.~Lombardi$^j$,
P.~Lombardi$^k$,
L.~Ludhova$^k$,
G.~Lukyanchenko$^{h}$,
I.~Machulin$^{h,i}$,
S.~Manecki$^q$,
W.~Maneschg$^{e}$,
S.~Marcocci$^u$,
E.~Meroni$^k$,
M.~Meyer$^d$,
L.~Miramonti$^k$,
M.~Misiaszek$^{g,j}$,
M.~Montuschi$^r$,
P.~Mosteiro$^n$,
V.~Muratova$^m$,
B.~Neumair$^s$,
L.~Oberauer$^s$,
M.~Obolensky$^{a}$,
F.~Ortica$^{l}$,
K.~Otis$^{p}$,
M.~Pallavicini$^c$,
L.~Papp$^s$,
L.~Perasso$^c$,
A.~Pocar$^{p}$,
G.~Ranucci$^k$,
A.~Razeto$^j$,
A.~Re$^k$,
A.~Romani$^{l}$,
R.~Roncin$^{j,a}$,
N.~Rossi$^j$,
S.~Sch\"onert$^s$,
D.~Semenov$^m$,
H.~Simgen$^{e}$,
M.~Skorokhvatov$^{h,i}$,
O.~Smirnov$^{b}$,
A.~Sotnikov$^{b}$,
S.~Sukhotin$^h$,
Y.~Suvorov$^x$,
R.~Tartaglia$^j$,
G.~Testera$^c$,
J.~Thurn$^w$,
M.~Toropova$^h$,
E.~Unzhakov$^m$,
A.~Vishneva$^{b}$,
R.B.~Vogelaar$^q$,
F.~von~Feilitzsch$^s$,
H.~Wang$^x$,
S.~Weinz$^z$,
J.~Winter$^z$,
M.~Wojcik$^{g}$,
M.~Wurm$^z$,
Z.~Yokley$^q$,
O.~Zaimidoroga$^{b}$,
S.~Zavatarelli$^c$,
K.~Zuber$^w$,
G.~Zuzel$^{g}$
}

\affiliation{  \vspace{10 pt} 
\address(a) \mbox{AstroParticule et Cosmologie, Universit\'e Paris Diderot,~CNRS/IN2P3,~CEA/IRFU,} 
\mbox{Observatoire de Paris, Sorbonne Paris Cit\'e, 75205 Paris Cedex 13, France} \\
\address(b) \mbox{Joint Institute for Nuclear Research, 141980 Dubna, Russia} \\
\address(c) \mbox{Dipartimento di Fisica, Universit\`a degli Studi e INFN, 16146 Genova, Italy} \\
\address(d) \mbox{Institut f\"ur Experimentalphysik, Universit\"at Hamburg, 22761 Hamburg, Germany} \\
\address(e) \mbox{Max-Planck-Institut f\"ur Kernphysik, 69117 Heidelberg, Germany} \\
\address(f) \mbox{Kiev Institute for Nuclear Research, 03680 Kiev, Ukraine} \\
\address(g) \mbox{M.~Smoluchowski Institute of Physics, Jagiellonian University, 30348 Krakow, Poland} \\
\address(h) \mbox{NRC Kurchatov Institute, 123182 Moscow, Russia} \\
\address(i) \mbox{National Research Nuclear University MEPhI (Moscow Engineering Physics Institute), 115409 Moscow, Russia} \\
\address(j) \mbox{INFN Laboratori Nazionali del Gran Sasso, 67010 Assergi (AQ), Italy} \\
\address(k) \mbox{Dipartimento di Fisica, Universit\`a degli Studi e INFN, 20133 Milano, Italy} \\
\address(l) \mbox{Dipartimento di Chimica, Biologia e Biotecnologie, Universit\`a e INFN, 06123 Perugia, Italy} \\
\address(m) \mbox{St. Petersburg Nuclear Physics Institute NRC Kurchatov Institute, 188350 Gatchina, Russia} \\
\address(n) \mbox{Physics Department, Princeton University, Princeton, NJ 08544, USA} \\
\address(o) \mbox{Chemical Engineering Department, Princeton University, Princeton, NJ 08544, USA} \\
\address(p) \mbox{Amherst Center for Fundamental Interactions and Physics Department,} \mbox{University of Massachusetts, Amherst, MA 01003, USA} \\ 
\address(q) \mbox{Physics Department, Virginia Polytechnic Institute and State University, Blacksburg, VA 24061, USA} \\
\address(r) \mbox{Dipartimento di Fisica e Scienze della Terra  Universit\`a degli Studi di Ferrara e INFN,  44122, Ferrara, Italy} \\
\address(s) \mbox{Physik-Department and Excellence Cluster Universe, Technische Universit\"at  M\"unchen, 85748 Garching, Germany} \\
\address(t) \mbox{Lomonosov Moscow State University Skobeltsyn Institute of Nuclear Physics, 119234 Moscow, Russia} \\
\address(u) \mbox{Gran Sasso Science Institute (INFN), 67100 \L'Aquila, Italy} \\
\address(v) \mbox{Department of Physics, University of Houston, Houston, TX 77204, USA} \\
\address(w) \mbox{Department of Physics, Technische Universit\"at Dresden, 01062 Dresden, Germany} \\
\address(x) \mbox{Physics and Astronomy Department, University of California Los Angeles (UCLA), Los Angeles, CA 90095, USA} \\
\address(y) \mbox{Department of Physics and Astronomy, University of Hawaii, Honolulu, HI 96822, USA} \\
\address(z) \mbox{Institute of Physics and Excellence Cluster PRISMA, Johannes Gutenberg-Universit\"at Mainz, 55099 Mainz, Germany} \\
\address(za)\mbox{Lomonosov Moscow State University Faculty of Physics, 119234 Moscow, Russia} \\
\address(zb)\mbox{Laboratorio Subterr\'aneo de Canfranc, Paseo de los Ayerbe S/N, 22880 Canfranc Estacion Huesca, Spain}
}

\collaboration{Borexino Collaboration}
\noaffiliation


\begin{abstract}
Borexino is a liquid scintillation detector located deep underground at the Laboratori Nazionali del Gran Sasso (LNGS, Italy). Thanks to the unmatched radio-purity of the scintillator, and to the well understood detector response at low energy, a new limit on the stability of the electron for decay into a neutrino and a single mono--energetic photon was obtained. This new bound, $\ensuremath{\tau\geq 6.6 \times10^{28}}$~yr at 90 \% C.L., is two orders of magnitude better than the previous limit.
 \end{abstract}


\maketitle

The conservation of electric charge, suggested since the 19th century, is fundamental to the physics of the Standard Model as a direct consequence of Maxwell's equations and the unbroken $U(1)$ gauge symmetry of the electro-weak theory. Despite the present undisputed validity of this law, experimental tests of charge conservation remain a way to search for physics beyond the Standard Model, and deserve to be investigated with the highest possible sensitivity. An experimental search for the hypothetical charge non-conserving decay of the electron, which is the lightest known charged particle, into a neutrino and a photon is reported in this paper. No presently viable theory predicts such a decay, and a large charge violation is excluded by the absence of macroscopic effects in matter. The electron decay is, however, discussed in the literature, e.g. \cite{bib:Okun,2,3,4,7,bib:Aharonov} and references therein.

The Borexino detector \cite{bib:dete} is a unique tool to undertake a search for electron decay. The unmatched radio-purity of the liquid scintillator (a mixture of PC and PPO \cite{bib:liquid}) was obtained by means of a very careful selection and cleaning of materials, the development of innovative purification techniques~\cite{bib:bx1,bib:bx2}, and the extreme care taken during procurement and handling of the scintillator fluid and the filling of the detector. In addition, the 
successful measurement of several solar neutrino components with Borexino
($^7$Be~\cite{Be711,Be7Long}, $^8$B~\cite{bib:B8}, pep~\cite{bib:pep}, and pp~\cite{PP14}) demonstrates an unprecedented sensitivity which can be 
applied to the search for a two-body $e\rightarrow \nu_e + \gamma$ decay resulting in the emission of a 256 keV photon.  

The Borexino collaboration has already searched for this decay using the Counting Test Facility (CTF)~\cite{CTF}, a 4 m$^3$ liquid scintillator prototype detector operated to prove the feasibility of Borexino and to develop the necessary purification techniques for the scintillator. This earlier measurement used PXE scintillator~\cite{PXE08}, and set a lower limit for the lifetime of this electron decay mode of $4.6\times10^{26}$~yr~\cite{eDecayCTF}.
The work reported here uses a mixture of PC and PPO scintillating solution in the Borexino detector instead.
A sensitivity improvement of more than two orders of magnitude is obtained thanks to a larger fiducial mass, higher statistics, lower background, and significantly improved data analysis. The latter is the result of a deeper understanding of the detector energy response, obtained by extensive calibration campaigns \cite{Calib12} and encoded in a tuned Geant4 simulation.
 
In Borexino, neutrino interactions and background events are detected by means of scintillation light collected by some 2212 photomultipliers (PMTs) (various PMTs were taken offline over the years of Borexino operations due to electronic failure). The PMTs are supported by a Stainless Steel Sphere (SSS). The sphere contains, a thin concentric nylon vessel which separates the detection scintillator fluid, a PC+PPO(1.5 g/l) scintillating mixture, from an outer buffer liquid, a mixture of PC and DMP without PPO. A second larger nylon vessel acts as a Radon barrier. The SSS is contained in a large Water Tank (WT) instrumented with PMTs, shielding the inner spheres against external radiation and providing an active muon Cherenkov veto. For more details about the Borexino detector see \cite{bib:dete} and \cite{bib:muon}. 

The light collected by all the PMTs provides four basic measurements: 1) the total deposited energy, reconstructed from the number of collected photo--electrons; 2) the event position, reconstructed by means of a time--of--flight fit between the PMT signals; 3) the particle type ($\alpha$--like or $\beta$--like or muon), reconstructed from pulse shape analysis; and 4) the WT Cherenkov signal. For more details see \cite{Be7Long}. 

The analysis performed for this paper has taken advantage of the recent measurement of {\it pp} solar neutrinos~\cite{PP14}, as the expected photon visible energy from the hypothetical two-body electron decay occurs in the region where the {\it pp} neutrino signal is dominant. The correlation between {\it pp} neutrinos and the electron decay signal is discussed below. 
The study of the lower energy region of the spectrum (165-590 keV) poses special challenges related to the linearity of the energy response, light quenching processes, and $^{14}$C event pile--up. These were tackled for the measurement of {\it pp} neutrinos for which special tools were developed, which are also applied in this work.

Data used in this search were collected during the ``Phase 2'' operations of the Borexino experiment, started in 2012 after a set of calibration runs and more than a year of liquid scintillator purification. 
The calibration was performed with internal $\gamma$, $\beta$, $\alpha$, and neutron sources, which yielded a meticulous understanding of the detector response over a large energy range. Scintillator purification, done by means of water extraction and nitrogen stripping, substantially reduced radioactive backgrounds. In particular, $^{85}$Kr concentration is now compatible with zero (from $\sim$35 cpd/100 t in Phase 1), and the $^{210}$Bi content was reduced by a factor $\sim$4, to about 20 cpd/100 t. $^{238}$U and $^{232}$Th concentrations were at a record low, $\le 10^{-19}$ g/g. Data for this work were acquired from January 2012 to May 2013, corresponding to 408 live days.

\begin{figure}[t]
\begin{centering}
\includegraphics[width=0.53\textwidth]{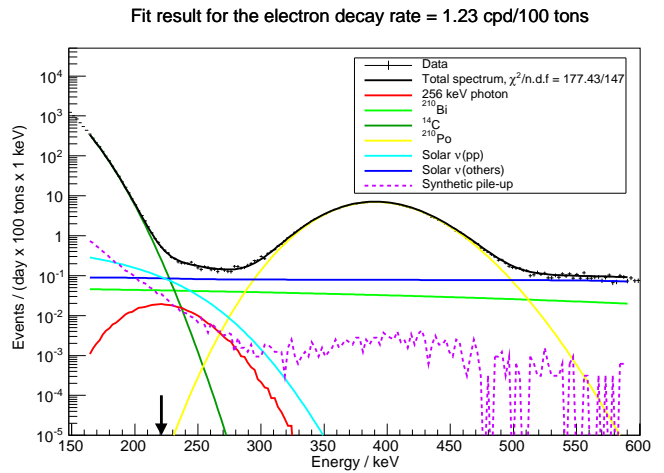}
\par\end{centering}
\caption{\label{fig1} Energy spectrum between 150 and 600 keV. The most prominent features
are the $^{14}$C $\beta$-spectrum (green), the peak at about 400 keV from $^{210}$Po 
$\alpha$-decays, and the solar neutrinos, grouped in the blue curve except the
crucial pp neutrinos, which are shown in cyan. The effect of event pile-up, mostly overlapping $^{14}$C events, is shown in dashed-pink.  The hypothetical mono--energetic 256
keV $\gamma$ line is shown in red at its 90\% exclusion C.L. with an arrow indicating
the mean value of the detected energy, which is lower than 256 keV because of quenching. {\mbox {The fit is done in the range 164-590 keV.}}}
\end{figure}

The energy spectrum used for the electron decay search is shown in Fig. \ref{fig1} (black points with error bars) together with the main fitted components (color lines).  At low energy (below 200 keV) the count rate is dominated by the $\beta$-decays of $^{14}$C,
with a measured abundance of $(2.7\pm0.1)\times10^{-18}$ g/g \cite{PP14} with respect to $^{12}$C. The mono-energetic peak in the central part of the spectrum corresponds to 5.3 MeV $\alpha-$particles from $^{210}$Po decay, which shifts downward to approximately 400 keV electron equivalent by quenching in the scintillator\cite{bib:birks}. The arrow indicates the position of the hypothetical 256 keV $\gamma$ peak from electron decay.

The event rate from $^{14}$C, while intrinsically low, still yields large event statistics over the entire scintillator volume, on the order of $5\times10^{5}$ events for the lowest energy bin used in this analysis. This requires very precise fitting models to keep the systematic uncertainties at or below the statistical fluctuations of approximately 0.14\%. A consequence of $^{14}$C decays is a non-negligible occurrence of pile-up events, when two or more independent decays (mostly $^{14}$C) occur close enough in time not to be separated. Two events can be distinguished with $>$50\% efficiency when they are separated by more than 230 ns. The energy spectrum of pile-up events in the region of interest above the $^{14}$C end--point is similar to the electron recoil spectrum induced by {\it pp} solar neutrinos.  
The pile--up spectrum obtained for the measurement of {\it pp} neutrinos~\cite{PP14}, shown in magenta in Fig.\ref{fig1}, is included as a separate component in the spectral fit used in this analysis and labeled ``synthetic'' pile--up.
The synthetic pile--up spectrum is constructed by overlapping triggered events with PMT hits recorded in the tail of the acquisition gate of such events, well after the triggered scintillation pulse has decayed away. These late PMT hits represent a data--driven, threshold--less and random sample of activity (data $+$ noise) in the detector. Added to triggered events they boost the pile--up contribution by a known amount, allowing the pile--up spectrum to be extracted.

The scintillation signal from a 256 keV $\gamma$ produced in the \ensuremath{e^-\rightarrow\gamma+\nu_e} decay is equivalent to one produced by a $220\pm0.4$~keV electron \cite{Calib12}. The energy shift is due to the partial light loss from quenching, i.e. the non--linearity of the scintillation response with electron energy. Quenching is modeled by the standard Birks' formalism~\cite{bib:birks}, which relates the density of light production $\frac{dL}{dx}$
to the ionisation density $\frac{dE}{dx}$:
\begin{equation}
\frac{dL}{dx}\sim\frac{\frac{dE}{dx}}{1+kB\frac{dE}{dx}},\label{Birks}
\end{equation}
where $k$ and $B$ are the Birks' parameters. The average number of collected photo--electrons (p.e.) $Q$ produced by an electron of energy $E$ can be obtained by integrating (\ref{Birks}). It is convenient to present the result in the form:
\begin{equation}
Q=LY\cdot E\cdot f(kB,E),\label{Light}
\end{equation}
where $LY$ is the light yield for electrons expressed in p.e.$\cdot$MeV$^{-1}$ and $f(kB,E)$ is a light deficit function, i.e. the result of the integration
of eq. \ref{Birks} along the path normalised to unity at 1 MeV. The $f(kB,E)$ is a monotonically increasing function in the region of interest. Thus, the average light yield of a $\gamma$ absorbed by multiple Compton scatterings at low energy followed by photo--absorption, is lower than that released by a single electron of the same initial energy. This fact is crucial because the quenched 256 keV--$\gamma$ energy partially overlaps with the $^{14}$C tail, which then required a special analysis. 

The number $n$ of PMTs that give a valid hit within a time window of 230 ns is approximately proportional to the energy deposit and therefore to the total charge Q collected by the PMTs. {\mbox {The relation between $n$ and $Q$ is:}}
\begin{equation}
Q=-\frac{N_{PMT}\log(1-\frac{n}{N_{PMT}})}{1+g_c\log(1-\frac{n}{N_{PMT}})},\label{eq:Charge}
\end{equation}
where $\text{N}_{\text{PMT}}$ is the total number of PMTs of the detector, and $g_c$ is a geometric correction factor obtained by means of MC. 
This approach is the same used in solar neutrino analysis and is described in detail in \cite{Be7Long} and \cite{PP14}. The statistics of $n$ is easier to model with respect to the number of collected photoelectrons $Q$ because the statistical distribution of multiple hits in PMTs depends on the details of the electronic response, and it is not known with sufficient precision.

The most crucial part of the analysis is the behaviour of the energy resolution as a function of the energy. Thus the variance of the energy resolution (in terms of the used energy estimator) is modeled as \cite{bib:Pablo}: 
\begin{equation}
\sigma_{n}^{2}=N(p_{0}-p_{1}v_{1})+n^{2}(v_{T}(n)+v_{f}(N))+\sigma_{d}^{2}+\sigma_{int}^{2}
\end{equation}
where $p_0=1-p_1$, $N=\left<f(t)\right>_{T}$ is the average number of operating PMTs during the period of the data acquisition, and $T$, $f(t)$ is the number of operating PMTs as a function of time, normalised as $f(0)=N_{0}$. Here $N_0$ is the number of working PMTs at the beginning, $v_{f}(N)=\left<f^{2}(t)\right>_{T}-\left<f(t)\right>_{T}^{2}$ is the variance of $f(t)$ over the period of data acquisition, $\sigma_{d}$ is the contribution of the dark noise (fixed at the measured value as an average over all PMTs) and, $\sigma_{int}$ is the contribution due to the smearing of the intrinsic line.

The probability $p_{1}$ that a single PMT is hit in a given event depends on event position, and is related to the energy estimator, $n$, as $n=N\cdot p_{1}$. The parameter, $v_{1}\equiv\frac{1}{\overline{p}_{1}}\left<p_{1}(\overrightarrow{r})v_{1}(\overrightarrow{r})\right>_{V}$, is the weighted average variance of the single PMT response, where $v_{1}(\overrightarrow{r})$
is defined as the variance of the $p_{1}$ (averaged over all PMTs) for an event whose position is $\overrightarrow{r}$. The quantity, $v_{1}$, is calculated over the detector's Fiducial Volume (FV) using Monte Carlo (MC) simulations. Due to the very narrow energy region of interest, - essentially the region of the $^{14}$C tail -, the energy dependence of $v_{1}$ in this analysis can be neglected. From the above the value of $v_{1}=0.17$ is computed. 

Finally, $v_{T}$ is the variance of the number of triggered PMTs over the FV for a  fixed energy.  It takes into account the non-uniform light collection over the detector's volume and the additional variability of the number of triggered PMTs in different locations within the FV. It is energy-dependent and was found to be proportional to energy in the region of interest. Its energy dependence was modelled as $v_{T}=v_{T}^{0}n$ \cite{Be7Long}, leaving the constant of proportionality as a free parameter of the fit.

The shape of the scintillation line (i.e., the energy response of the detector to a mono-energetic source uniformly distributed within FV) is another important component of this analysis. The familiar gaussian approximation fails to describe the tails of MC-generated mono--energetic peaks, even for the statistics on the order of $10^{3}$ events. In the previous $^{7}$Be solar neutrino analysis \cite{Be711} this problem was solved by using a generalised $\Gamma$--function~\cite{SM07} to fit the mono--energetic $^{210}$Po peak. However, while the quality of the fit to the $^{210}$Po peak is insensitive to the residual deviations in the tails, this is not the case for the $^{14}$C spectrum as all the events in the fraction of the $^{14}$C spectrum above the $^{14}$C end--point originate from spectral smearing so that the statistics in the tails are much higher. For these reasons, a different procedure was adopted. 

The ideal detector response to a point-like mono--energetic source in the center is an exact binomial distribution which can be well approximated by a Poisson distribution. However, the ideal width of the distribution must be modified to include the additional spreading of the signal due to various factors. The problem, both with binomial distribution and with its Poisson approximation, is that their width is defined by the mean value, which introduces an unwanted and unphysical correlation between the position of the peak and its width. To attack this problem, the response function was approximated by a Scaled Poisson Distribution (SPD) defined~by:
\begin{equation}
f(x)=\frac{\mu^{s x}}{(s x)!}e^{-\mu} \label{ScaledPoisson}
\end{equation}
where $x$ is the independent variable whose mean value is $n$ and variance $\sigma_n$. This function has two free parameters $\mu$ and $s$, which can be evaluated using an expected mean and variance. The agreement of this approximation with the detector response function was tested with the Borexino MC model. It was found at low energies that the function  (\ref{ScaledPoisson}) reproduces the scintillation line shape much better than a generalised $\Gamma$--function (GGF) up to statistics of $10^{8}$ events per bin, while at energies just above the $^{14}$C tail both distributions give comparable results. Therefore the SPD approximation was adopted, and the quality of the fit was estimated using a $\chi^{2}$ criterion. As an example, with $10^{7}$ mono--energetic events for $n$=50 (approximately 140 keV) we found $\chi^{2}/n.d.f.$=88.0/61 for the GGF compared to $\chi^{2}/n.d.f.$=59.3/61 for the SPD. In this example the events were uniformly distributed in the detector before the FV is selected.

As proven by MC calculations, the SPD works well in the region of interest despite the additional smearing due to the aforementioned factors. This is a result of the folding of the relatively narrow non-statistical distributions by the much wider base function. The MC shows that such an absorption results in the smearing of the total distribution without changing its shape. 

Only a fraction of the total response for the mono--energetic 256 keV $\gamma$ enters into the analysis window above threshold, which makes the signal look similar to the pp-spectrum and produces a strong correlation between them (see Fig. 1). Note the correlation between electron decay and pp solar neutrinos: with an unconstrained pp flux, 12  cpd/100 t of electron decay candidates correspond to about 134 cpd/100 t from pp solar neutrinos in the fit. To break the degeneracy, the pp-neutrino rate can be constrained either by the value measured by experiments other than Borexino or at that predicted by LMA-MSW theory. We chose to use data from radiochemical experiments only to obtain a model-independent result.

The deviation from the observed pp-neutrinos rate is controlled by a penalty term added to the $\chi^{2}$ in the form:
$$
\chi_{pp}^{2}=\frac{(pp-\left<pp\right>)^{2}}{\sigma_{pp}^{2}+\sigma_{FV}^{2}}
$$
where $pp$ is the pp-neutrino count rate found by the fit, $\left<pp\right>$=134~cpd/100 t is the rate expected in Borexino, and $\sigma_{pp}$=13.3~cpd/100 t its variance. These are calculated using the solar pp-neutrino flux obtained from the combined analysis of the GALLEX/GNO and SAGE experiments \cite{Radiochemical}. The parameter $\sigma_{FV}$=2.7 takes into account the systematic error (2\% as reported in \cite{PP14}) due to the uncertainty in the FV mass. 

A search for the 256 keV $\gamma$ line was then undertaken using 
``standard'' fit conditions, defined by:
a) an energy estimator: number of triggered PMTs in a fixed time window of 230 ns (npmts); b) a fit range: 62--220 npmts, corresponding to 164-590 keV;
c) a fiducial mass: 75.5 t (R$<$3.02~m and $|Z|<$1.67~m) which is the same as in solar $^{7}$Be analysis. The values of solar neutrino rates other than pp are constrained either at the results found by Borexino in a different energy range of R($^7$Be)=48$\pm$2.3~cpd/100 t \cite{Be711}, or fixed at the predictions of the SSM in the MSW/LMA oscillation scenario, R(pep)= 2.80~cpd/100~t, R(CNO)=5.36~cpd/100~t.
The $^{14}$C rate was constrained at the value found in an independent measurement R($^{14}$C)=40$\pm$1 Bq (or R($^{14}$C)=(3.456$\pm$0.0864)$\times$10$^{6}$~cpd/100 t). The pile-up rate was constrained at the values found with the above algorithm to be $321\pm7$~cpd/100 t. Other background components were left free ($^{85}$Kr,$^{210}$Bi and $^{210}$Po) and the rate of $^{214}$Pb R($^{214}$Pb)=0.06~cpd/100 t was calculated by means of identified $^{222}$Rn events, which were measured by the detection of $^{214}$Bi--$^{214}$Po coincidences. The light yield and the two energy resolution parameters ($v_{T}$ and $\sigma_{int}$) are left free in the fit. The position of the $^{210}$Po peak is also left free, and decoupled from the energy scale because of its poorly known quenching factor.

The detection efficiency of 256 keV $\gamma$s was determined by means of the MC code. A set of uniformly distributed $\gamma$s were simulated inside the entire IV. The events passing the same set of cuts used for real data selection and reconstructed within the FV were used to determine a global efficiency of $\epsilon$ = 0.264,
which includes FV cut.

The upper limit S for the given confidence level C.L. comes from the
following integral:

\begin{equation}
\int_{0}^{S}f(\xi)\mathrm{d}\xi=\mathrm{C.L.}\int_{0}^{\infty}f(\xi)\mathrm{d}\xi,\label{eq:int}
\end{equation}
where $f(\xi)$ is the probability distribution function. With this procedure, we found the 90\% C.L. upper limit S=379 events, in a period T=408 d and fiducial mass of 75.5~t. 

The corresponding lifetime $\tau_{BX}\geq7.2\times10^{28}$~yr (statistical only) was obtained using the relation:
\begin{equation}
\tau_{BX}\geq\frac{\epsilon N_{e}T}{S} 
\label{eq:time}
\end{equation}
where  $N_{e}=9.19\times10^{31}$ are the electrons in the IV mass of 278 t and $\epsilon$=0.264 is the detection efficiency.

The systematic error depends on three factors: the choice of the energy estimator, the value of the quenching parameters, and the knowledge of the fiducial volume. 

The main one is the choice of the energy estimator, and in particular the length of the 230 ns time window, which affects the relevance of pile--up events and dark noise hits. We have throughly studied this effect by means of MC simulations. It amounts, for this analysis, to about 8\%. 

The second one is the uncertainty is the position of the 256 keV $\gamma$--peak with respect to the $^{14}$C spectrum depends on the quenching factor, $kB$, and the precise position of the end-point of the $^{14}$C $\beta$--spectrum. Since the underlying $^{14}$C $\beta$--spectrum is falling exponentially, a shift of the peak position to lower energy decreases the sensitivity of the search and vice versa.  In the present analysis the $kB$ parameter is fixed at $kB=0.0109$ $\text{cm}\cdot\text{MeV}^{-1}$, determined from calibration data with mono--energetic $\gamma$ sources. In order to study the related systematics in a model independent way, the position of the $\gamma$-peak within the bounds was obtained by calibration (1\%) instead of varying the $kB$ value. It should be noted that no degradation of the fit quality for a wide range of $kB$ values was observed, due to the absence of known $\gamma$-sources in the region of interest and, therefore, in the fit model. The final result is that $kB$ error adds less than 2\% to the total systematic error. 

The third systematic error is the knowledge of the FV mass which, at the energies of interest, adds an uncertainty of less than 1\% on electron decay rate.

The three effects described above are correlated. The total systematic error was obtained by building a set of probability profiles for a corresponding set of input parameters. The probability profiles were renormalised to physical regions, excluding non--physical values in case of a negative number of candidates where the region below zero was cut and the p.d.f. was renormalised to unity. The final probability profile was obtained as the weighted sum of the separate p.d.f.s, with the weights being the probability of occurrence of the corresponding value of the parameter. The normal distribution was used for both the error on the FV and for the relative shift of the 256 keV
$\gamma$ with respect to the electrons scale. The addition of the total systematic error degrades the purely statistical result by about 8\%.

The lower limit on the mean lifetime for the decay $e^{-}\rightarrow \gamma + \nu$ is $\tau_{BX}\geq6.6\times10^{28}$~yr (90\% C.L.), which improves the existing limit from CTF~\cite{eDecayCTF} by more than two orders of magnitude. The sensitivity is such that a 5$\sigma$ discovery signal would have been possibile with an electron lifetime of $1.9\times10^{28}$ yr.

The Borexino program is made possible by funding from INFN (Italy), NSF (USA), BMBF, DFG, and MPG (Germany), RFBR: Grants 14-22-03031,15-02-02117 and 13-02-12140, RFBR-ASPERA-13-02-92440 (Russia), and NCN Poland (UMO-2012/06/M/ST2/00426).  We acknowledge the generous support and hospitality of the Laboratori Nazionali del Gran Sasso (LNGS).

\end{document}